# Modulating Peierls distortion of 1T' MoS$_2$ via charge doping: a new charge density wave phase, reversible phase transition, and excellent electromechanical properties

Kaiyun Chen[1], Junkai Deng[1, *], Qian Shi[1], Xiangdong Ding[1], Jun Sun[1], Sen Yang[1, *] and Jefferson Zhe Liu[2, *]

[1] MOE Key Laboratory for Nonequilibrium Synthesis and Modulation of Condensed Matter, State Key Laboratory for Mechanical Behavior of Materials, Xi'an Jiaotong University, Xi'an 710049, China

[2] Department of Mechanical Engineering, The University of Melbourne, Parkville, VIC 3010, Australia

*junkai.deng@mail.xjtu.edu.cn; *yangsen@mail.xjtu.edu.cn; *zhe.liu@unimelb.edu.au

## Abstract

The 1T' phase of transition metal dichalcogenides (TMDs) is a low symmetry charge density wave (CDW) phase, which can be regarded as a periodically distorted structure (Peierls distortion) of the high symmetry 1T phase. In this paper, using density functional theory (DFT) calculations, we report that the positive charge (hole) injection is an effective method to modulate the Peierls distortion of MoS$_2$ 1T' for new CDW phase and superior electromechanical properties. A new stable CDW phase is discovered at hole doping level of ~ 0.10h$^+$/atom, named as 1T$_t$'. The hole charging and discharging can induce a reversible phase transition of MoS$_2$ among the three phases, 1T, 1T' and 1T$_t$'. Such reversible phase transition leads to superior electromechanical properties including a strain output as high as -5.8% with a small hysteresis loop, multi-step super-elasticity, and multi-shape memory effect, which are valuable in active electromechanical device designs at nanoscale. In-depth analysis on the change of electronic structure under hole doping was performed to understand the new CDW phase and the observed phase transition. Using charge doping to modulate the Peierls distortion in two-dimensional materials can serve as a general concept for nano-active material designs.

# I. INTRODUCTION

The two dimensional (2D) transition metal dichalcogenides (TMDs), which has a sandwich crystal structure with M (Transitional element) atoms in the middle and X (S, Se or Te) atoms on the top and bottom layers, have fueled extensive research efforts for excellent electronic, magnetic and chemical properties.[1-3] They have great potentials for next-generation electronics, optoelectronics, and electromechanical devices.[4-5] The abundant physical and chemical properties of TMD family can be attributed to not only their chemical diversity but also their multiple stable crystal structures. The 2H phase often is a direct band gap semiconductor with intrinsic in-plane electric polarization.[6] It is recognized as a good candidate for field effect transistor, valley-based nonvolatile random access memory, and valley filter devices.[7-10] The 1T phase is metallic and often shows excellent catalytic properties for hydrogen evolution reaction.[11-12] The low symmetry crystal phases, such as 1T' and 1T'', are attracting increasing attention in the past few years. They can be regarded as a periodically distorted crystal structure from the high symmetry 1T phase.[13-14] The periodic lattice distortion in 1T' or 1T'' phase induces periodic modulation of charge density, leading to the charge density wave (CDW) state. Recent researches show that the CDW state embodies a great opportunity for various applications such as catalysis, energy storage, superconductivity, and surface chemistry.[2, 15-19]

Peierls distortion is a primary mechanism for the formation of the 1T' phase (and then the CDW state).[13] Peierls instability is originally proposed to explain the spontaneous periodic lattice distortion of one dimensional conductor.[20] A superstructure of a crystalline material has a reduced First Brillouin zone (BZ). Its electronic band structure can be formed by folding the band structure of a unit cell into the reduced BZ. For some metallic materials, the band folding will cause the Fermi level ($E_f$) of some band exactly at one zone boundary, leading to the so-called Peierls instability. This will promote a spontaneous lattice distortion to open a small band gap at the zone boundary. The lowing of the electronic energy levels below the $E_f$ will reduce the total energy of system, compensating the strain energy arising from the lattice distortion and thus stabilizing the distorted superstructure. For some 2D TMDs, the periodic lattice distortion will generate dimerization between two adjacent lines of metal atoms along a high symmetry zig-zag direction, forming parallel zig-zag chains of bonded M atoms. It is clear that the Peierls distortion is an intrinsic coupling between electronic structure and lattice distortion/deformation. Therefore, it is reasonable to propose that tuning the electronic structure of 1T' TMD could lead to significant lattice distortion/deformation and thus superior electromechanical properties.

It is well recognized that miniaturization of electromechanical device will bring a revolution to humanity in the coming decades, synonymous with the efforts of miniaturizing electronic device in those previous.[21-24] They promise and deliver a myriad of applications within industry and biomedicine. An electromechanical actuator – a device that converts electrical energy to mechanical motion – is the core component of many such devices. Consequently, research interrogating micro- and nano-actuators becomes increasingly essential. But conventional actuation materials, such as piezoelectric and shape memory alloys, exhibit significantly compromised and even diminished performance at small length scale. The concept of modulating Peierls distortion in 1T'-TMD for superior electromechanical properties could enable the design of various high-performance electromechanical actuators at nanoscale.

In this paper, we employed density functional theory (DFT) calculations to explore this concept in $MoS_2$, the most well-known TMD material. Our results show that hole doping can significantly change the lattice structures of $MoS_2$. It stabilizes the 1T phase and more interestingly a new CDW phase $1T_t'$. Tuning the hole doping level can switch the energetic order of these three phases, leading to reversible multi-phase transition. The electromechanical strain output can be as high as 5.8%. The multi-phase transition also induces novel multi-step super-elasticity and multi-shape memory effect.

## II. COMPUMTATIONAL METHODS

Our DFT calculations were conducted by using the Vienna Ab-inito Simulation Package (VASP).[25] The Perdew-Burke-Ernzerhof (PBE)[26] exchange-correlation functional and projector augmented wave[27] method were adopted. The 2D $MoS_2$ was placed in *x-y* plane. A vacuum layer of certain thickness was placed in *z*-direction in our supercells. Periodic boundary condition was applied in all three directions. The cutoff energy was set as 400 eV. The Monkhorst−Pack $k$-point mesh[28] density of 21 × 19 × 1 was used for 1T'-MoS$_2$ primitive unit cell and 1 × √3 × 1 supercell of 1T-MoS$_2$. The atomic positions and lattice constants were fully relaxed until total energy difference was less than $10^{-4}$ eV, and the force was less than 0.01 eV/Å. A higher convergence criterion of $10^{-6}$ eV was adopted to calculate electronic structures.

In VASP calculations, the injected charges (electrons and holes) were compensated by a background jellium to maintain the charge neutrality in the supercell. The supercell size in $z$-direction affects the background jellium charge density and could affect simulation results, particularly the relaxed lattice constants. We tested the $z$-direction size of 15Å. Fig. S1 summarizes the lattice constant $a$ and $b$ of 1T'-MoS$_2$ under different hole injection, which were obtained using automatic relaxation in VASP calculations. The lattice constants decrease with the increase of hole injection. There is an abrupt change under Q ~ 0.10 – 0.12 h$^+$/atom, corresponding to 1T'-1T phase transition. To confirm, we did an independent investigation. Some in-plane strains (from -2% to 2%) were applied on the lattice constants of 1T'-MoS$_2$. Then a set of hole injection was applied to the supercells. The total energy values of these charged MoS$_2$ slabs was calculated by using the correction method proposed by Reed et. al.[29] At given hole doping, the relationships between lattice constant strain and the corrected total energy are shown in Fig. S2. The equilibrium lattice constants decrease with increasing hole injection. This is consistent to the trend of lattice constant change in the VASP automatic relaxation calculations. Taking the zero charged 1T'-MoS2 as reference, the lattice strains along $y$ direction $\varepsilon_y$ from the automatic relaxation calculation are about -0.3%, -0.6%, -1.0% and -1.7% under Q = 0.02, 0.04, 0.06 and 0.08h$^+$/atom, respectively. The lattice strains $\varepsilon_y$ from the correction method are about -0.5%, -0.7%, -1.5% and 2.0% under Q = 0.02, 0.04, 0.06 and 0.08h$^+$/atom, respectively. The good agreement suggests that $z$ =15Å is appropriate for hole injection. In addition, in Fig. 2a, the corrected total energy value of 1T and 1T' has a cross point at Q = 0.10h$^+$/atom. It is consistent to the phase transition point during automatic relaxation of supercell with $z$ =15Å in VASP calculations. For convenience, supercells with $z$-direction size of 15Å were chosen for most of DFT calculations.

## III. RESULTS AND DISCUSSIONS

### A. Hole doping induced phase transition between 1T' and 1T MoS$_2$.

The crystal structures of 1T and 1T' phase of MoS$_2$ are depicted in Fig.1a. 1T-MoS$_2$ is a high symmetry phase with P$\bar{3}$m2 space group. It has a rhombohedral primitive unit cell. The Mo-Mo interatomic distances between two adjacent columns (along $x$ crystalline direction) are equal. The 1T' phase has a rectangle unit cell corresponding to the $1 \times \sqrt{3}$ supercell of the 1T phase (the blue rectangle in Fig. 1). Compared with the 1T phase, the two adjacent Mo atom columns in 1T' phase move toward each other in the $y$ direction to form dimers. The Mo-Mo dimer bond can be observed from the partial charge density in Fig. S3. Fig. 1a (right figure) shows clear zig-zag Mo dimer chains. Such a periodic lattice distortion leads to changes of interatomic distances and lattice constants. Along the $y$ direction, the Mo-Mo interatomic distances show periodic variation, *i.e.* $D_{Mo1\text{-}Mo2}/D_{Mo2\text{-}Mo3}$ = 0.73. The rectangle unit cell of 1T and 1T' have lattice constants $a$ = 3.186Å, $b$ = 5.520Å, and $a$ = 3.178Å, $b$ = 5.713Å, respectively. Taking the 1T' phase as a reference, the lattice mismatch strain is 0.3% and -3.5% along $x$ and $y$ direction, respectively.

DFT calculations showed that the 1T' phase is more stable than the 1T phase.[30] The energy barrier to separate the 1T phase from stable 1T' phase is nearly zero, suggesting a spontaneous phase transition. To understand the physical origin of the phase transition, we calculated the electronic band structures. The results are shown in Fig. 1b and c. The band structure of the 1T phase corresponds to the $1 \times \sqrt{3}$ rectangle supercell (Fig. 1b), which is comparable to the 1T' unit cell (Fig. 1c). In Fig. 1b, there is a band (mainly $d_{x^2-y^2}$ orbital) continuously crossing the Fermi level exactly at the X-point, indicating the Peierls instability in the 1T phase. After a phase transition to 1T', there is a clear band gap opening at X-point in Fig. 1c. The downward moving of the energy levels at X-point lowers system total energy, compensating the lattice distortion strain energy and stabilizing the 1T' crystal structure. Our results confirm the Peierls distortion as the physical origin of the 1T to 1T' phase transition. Note that our results also reveal the quasi-one-dimensional Peierls distortion feature. For 2D and 3D systems, the Peierls distortion also manifests as the Fermi surface nesting. To further confirm, the Fermi surface of 1T MoS$_2$ primitive cell is shown in Fig. S4. Indeed, there is a nesting vector $q_M$ connecting the electron (red) and hole (blue) pockets.

In light of the intrinsic coupling between electronic structure and lattice distortion in 1T and 1T'-MoS$_2$, we postulate that charge doping would lead to lattice structure change or phase transition. The total energy values of 1T' and 1T phase (Fig. 2a) under different hole injection (0.0 - 0.2 h$^+$/atom) were calculated using the correction method proposed by Reed et al.[29] A cross point is at Q = 0.10h$^+$/atom in Fig. 2a. The 1T' phase is more stable than 1T at a low hole doping level (Q < 0.10h$^+$/atom), whereas 1T is more stable under a relatively high hole doping (Q > 0.10h$^+$/atom). This result indicates that hole injection/extraction should induce a reversible phase transition. To get detailed information of structure changes, we adopted a successive way of hole injection/extraction in further DFT calculations. In other words, the fully relaxed structure of a hole doping level was taken as the initial structure of the following slightly enhanced or reduced hole doping level in VASP calculations. Fig. 2b, 2c and 2d summarize the variation of crystal structures upon hole injection/extraction. With hole injection up to Q = 0.10h$^+$/atom, the lattice constants $a$ and $b$ gradually reduce from $a$ = 3.178Å, $b$ = 5.713Å to $a$ = 3.104Å, $b$ = 5.376Å. Taking the neutral 1T' as a reference state, the strain is up to -3.2% and -3.4% along $x$ and $y$ direction, respectively. The $D_{Mo1-Mo2}/D_{Mo2-Mo3}$ ratio increases from 0.73 to 0.78. Beyond 0.10h$^+$/atom, the $D_{Mo1-Mo2}/D_{Mo2-Mo3}$ ratio jumps to 1.0, indicating a phase transition to 1T phase. The lattice constant $a$ exhibits an abrupt increase, whereas lattice constant $b$ shows a sharp drop. At Q = 0.11 h$^+$/atom, the lattice strain is $\varepsilon_x$ = -2.5% and $\varepsilon_y$ = -5.8%. These large electromechanical strain outputs arise from hole injection induced lattice contraction and the 1T' to 1T phase transition. Further increasing hole doping to 0.20 h$^+$/atom leads to lattice strain $\varepsilon_x$ = -3.7% and $\varepsilon_y$ = -7.2%.

In the reverse direction (hole extraction process), the lattice constants of the 1T phase gradually increase following the same curve of the hole injection. Surprisingly, the 1T does not transform to the 1T' at the identified critical point Q = 0.11h$^+$/atom. The phase transition takes place at a lower charging level of Q = 0.08h$^+$/atom. As a result, there is a hysteresis loop. A close inspection of Fig. 2b indicates that some structures in the hole extraction curve have $D_{Mo1-Mo2}/D_{Mo2-Mo3}$ ≈ 0.853 distinctive from both 1T and 1T' phases, suggesting a third stable phase of MoS$_2$. Indeed, further analysis confirms this new phase (named as 1T$_t$'), which will be discussed later.

To understand the physical origin of the observed hysteresis loop, we calculated total energy as a function of $y$ lattice constant under a given hole doping level of Q = 0.08, 0.09, 0.10, and 0.11h$^+$/atom, respectively. The results are summarized in Fig. 3. At a relative low hole injection of Q = 0.08h$^+$/atom, there is only one minimum energy point at $b$ = 5.625Å, corresponding to the 1T' phase. But two more distinctive curve segments at smaller lattice constants can be observed, suggesting two other possible structures or phases. Further examination of the crystal structures (*i.e.*, D$_{Mo1-Mo2}$/D$_{Mo2-Mo3}$ ≈ 1.0) suggests the first curve segment should belong to the 1T phase. The second one resembles 1T' but with a different D$_{Mo1-Mo2}$/D$_{Mo2-Mo3}$ value. We named this new phase as 1T$_t$'. The zero energy barriers (Fig. 3a) indicate these two phases unstable. Thus, in Fig. 2b, 2c, and 2d, only 1T' phase can be observed at Q=0.08h$^+$/atom. Increasing the hole injection slightly to Q=0.09h$^+$/atom, a non-zero energy barrier shows up to separate 1T$_t$' from 1T'. But 1T' is still the most stable phase. At Q = 0.10h$^+$/atom in Fig. 3c, all these three phases are stable, and the 1T$_t$' phase becomes the most stable phase. At Q = 0.11h$^+$/atom, the 1T phase becomes most stable. Our results conclude that hole doping will stabilize 1T and a new 1T$_t$' phase. Hole doping will alter the energetic orders of the three phases, leading to a reversible phase transition.

Fig. 3 also explains the origins of the observed hysteresis loop in Fig. 2. During the hole injection process, 1T' is the most stable phase up to Q = 0.09h$^+$/atom. At Q = 0.10h$^+$/atom, the 1T$_t$' becomes the most stable phase. But due to the energy barrier, MoS$_2$ remains at the 1T' phase. The energy barrier disappears until Q ~ 0.11h$^+$/atom in Fig. 3d. But the 1T' phase should directly transform to the 1T phase (instead of 1T$_t$'), as observed in Fig. 2b, 2c, and 2d. It is because the energy barrier separating 1T and 1T$_t$' also disappears, and the 1T phase is the most stable one. During hole extraction, in Fig. 3, the total energy of 1T move upward with respect to 1T$_t$' and 1T'. At Q = 0.10h$^+$/atom, the 1T$_t$' is most stable, and there is nearly zero energy barrier between 1T and 1T$_t$' (Fig. 3c). The 1T phase should therefore transform spontaneously to 1T$_t$' near 0.10h$^+$/atom, which is consistent to Fig. 2. At Q = 0.09h$^+$/atom, despite 1T' is the most stable, the existence of non-zero energy barrier will keep MoS$_2$ in 1T$_t$' phase. At Q = 0.08h$^+$/atom, 1T$_t$' transforms to 1T' as the energy barrier disappears. The energy barrier between these phases is the reason why we observe $1T_t$' (green star) near Q = 0.09 - 0.10h$^+$/atom in Fig. 2 and the hysteresis loop.

Here we should be aware that the hysteresis loop could be observed only at a low temperature. At a high temperature, thermal excitation could promote MoS$_2$ to overcome the energy barriers, and thus the loop tends to diminish. Nevertheless, two consecutive phase transition: from 1T' to $1T_t$' and then from $1T_t$' to 1T, should be observed during hole injection and vice versa during hole extraction.

The band structures of 1T, $1T_t$' and 1T' are calculated under 0.10h$^+$/atom hole injection, under which all these three phases are mechanically stable (Fig. 3c). The results are shown in Fig. 4. The band structures of 1T and 1T' phases resemble those at the charge neutral case. The notable difference is the downward shift of $E_f$. As such, for 1T, the $E_f$ of the band is not at the X point anymore. The driving force for the Peierls distortion in 1T-MoS$_2$ diminishes. This should be the reason why the 1T phase becomes metastable in Fig. 3c. Note that this will take place only under sufficiently high hole doping, consistent to the fact that the total energy difference between 1T and 1T' phase gradually reduces upon hole doping (Fig. 2a). Meanwhile, the $E_f$ downshift of the 1T' phase empties the $d_{xz}$ and $d_{z^2}$ band, which corresponds to the Mo-Mo dimer bonds observed from the partial charge density results in Fig. S5. The weakened "dimer" bonds tend to destabilize the 1T' phase, consistent to the relative energy changing trend in Fig. 3.

The interesting case is $1T_t$'. This new phase seems to be a transition phase between 1T and 1T' in light of the band structure. In Fig. 4b (Q = 0.10h$^+$/atom) and Fig. S6 (Q = 0.09h$^+$/atom), the band structure of $1T_t$' near $E_f$ resembles that of 1T phase. But the deep valance bands are more similar to those in 1T' phase. If we focus on the variation of the band in the vicinity of the Fermi level upon hole doping (Fig. 1b, 1c, Fig.4 and Fig. S6), this trend is more clear. In the zero doping case, at X-point, the 1T has the band with a convex shape, but 1T' has a concave shape. It is due to the band opening arising from the Peierls distortion. With the hole doping in 1T', the driving force for the Peierls distortion reduces. As a result, that band tends to restore the one of 1T phase. Indeed, at Q = 0.09h$^+$/atom, that band in $1T_t$' becomes much flatter than that in 1T'. At Q = 0.10h$^+$/atom, that band in $1T_t$' closely resembles the convex shape of that in 1T phase. In addition, the electron local function (ELF) results in Fig. S7 and S8 suggest that the bond character of $1T_t'$ is much close to 1T under higher level hole doping (Q = 0.10h$^+$/atom) while it is close to 1T' under lower doping level (Q = 0.09h$^+$/atom). Our electronic structure analysis suggests that $1T_t'$ could be a transition state between 1T and 1T'. $1T_t'$ phase could be a result of competition between the energy decrease by band gap opening and the energy increase by lattice strain distortion. At a certain hole doping level, these two factors might achieve a delicate balance, and thus a stable $1T_t'$ appears.

## B. Superelasticity and shape memory effect under hole doping.

In addition to the large electromechanical strain output of MoS$_2$ being subject to hole injection/extraction, there are some other interesting electromechanical properties. Superelasticity (SE) is a special phenomenon for shape memory materials, and originated from stress-induced phase transition, *e.g.*, the austenitic/martensitic phase transition in shape memory alloys.[31-32] SE has a broad range of applications as sensors, actuators, and energy storage devices.[33] Fig. 5 shows our DFT calculated stress-strain curves of MoS$_2$ under hole doping Q = 0.11h$^+$/atom. Two conditions are considered, strain control (Fig. 5b) and stress control (Fig. 5a). For the strain control case, we calculate the slope of energy-strain relation in Fig. 3d as stress and plot them as a function of strain. In both cases, the stress-strain curves show two plastic-like strain plateaus at certain stress levels. In Fig. 5a, the first plateau begins at ~2% and ends around ~2.5%, corresponding to phase transition from 1T to $1T_t'$. Then the second plastic plateau starts from 4% to 8%, corresponding to the $1T_t'$ phase transforms to 1T'. Note that the unloading curve does not overlap the loading one in the region where the phase transition takes place, leading to a hysteresis loop.

For the stress control case, we need to use common tangent lines in the energy-strain curve (Fig. 3d) to determine the critical stress at which phase transition occurs, which is shown in Fig. S10. The stress-strain curve is summarized in Fig. 5b. Two plateaus, corresponding to 1T to $1T_t'$ and $1T_t'$ to 1T' phase transition, are observed. The first plateau from 0.3% to 0.9% is the phase transition between 1T and $1T_t'$, while the second one from 2% to 3.7% corresponds to the phase transition from $1T_t'$ to 1T'. Note that the two plateau regions have two phases co-existed in mechanical equilibrium (corresponding to the common tangent in Fig. S10a).

One clear advantage of SE is the significant improvement of mechanical yield strain.[33] The maximum strain along *y* direction of neutral 1T'-MoS$_2$ is 9% (Fig. 5a). It is 19% at Q = 0.11h$^+$/atom, about double increase. The 1T' phase will return to the 1T phase after the stress is released. The hysteresis loop in the stress-strain curve is usually observed for SE materials. This stress-strain hysteresis loop could be useful for some nanoscale devices as energy absorber or damper.[34]

The observed two-step plateau in Fig. 5 arises from the three different phases under the appropriate hole doping level (Fig. 3). At some hole doping levels, *e.g.*, 0.12 h$^+$/atom, only two phases 1T and 1T' can be observed in the energy-lattice constant curve. As such there is only one plateau in the stress-strain curve (Fig. S9). The multi-plateau is not often observed in shape memory alloys. This multi-plateau related superelasticity is always related with larger reversible strain and shape memory.[35-38]

Shape memory effect (SME) is a phenomenon that some materials can recover their original shape from various temporary pseudo-plastic deformation under an appropriate external stimulus, such as heat, light, moisture or electric/magnetic field.[39-42] The underlying physical mechanism is the reversible phase transition among two or more stable phases. The reversible phase transition among 1T, 1T$_t$' and 1T' phase predicted in our DFT calculations could enable the multi SME. A strategic diagram is illustrated in Fig. 6 by taking analogy to the shape memory alloys (SMA).[43] A charge neutral 1T' MoS$_2$ phase with multiple domains can be regarded as the initial state. The previous study has suggested that mechanical strain can induce switch of the domains and the change of multi-domain structure due to the small energy barrier (~ 0.2eV/f.u.).[44] This will fix the material into various temporary shapes as a result of the purposely applied mechanical load. In the recovery process, holes can be injected into the temporarily shaped MoS$_2$ (1T' phase) to trigger a transform to the 1T phase (Q > 0.10h$^+$/atom). All domains disappear since the 1T-MoS$_2$ is a high symmetry phase, analogous to the high symmetry austenite phase in SMA. When holes are extracted, the MoS$_2$ will firstly transform to a multi-domain 1T$_t$' structure (Q ~ 0.08h$^+$/atom) and then to the initial state which is the charge neutral 1T' multi-domain structure. The appearance of the third phase, *i.e.*, 1T$_t$', in the recovery step, could enable the multi SME.[33, 39, 45-46]

## IV. CONCLUSION

In this paper, using DFT calculation, we investigate a design concept of two-dimensional electromechanical active/actuation materials, *i.e.*, modulating Peierls distortion via hole doping. The most widely studied MoS$_2$ in TMD family is taken as an example. We find that hole doping can stabilize the high symmetry 1T phase and generate a new stable CDW phase 1T$_t$'. It is because the hole doping shifts the $E_f$ level to weaken the driving force for Peierls distortion. Its competition with the strain energy from lattice distortion seems to stabilize the 1T and 1T$_t$' phases. By controlling the hole doping level, the energetic order of the three phases will be changed, and thus lead to the spontaneous phase transition. The resultant phase transitions enable diverse superior electromechanical actuation properties including a very large reversible strain output ~5.8%, a multi-plateau superelastic behavior, and potential multi-shape memory effect. Since Peierls distortion generally exists in a broad class of 2D TMD materials, our study demonstrates a general concept of design 2D high-performance TMD based actuation materials. Besides, the new 1T$_t$' phase observed in our DFT calculations also suggests a new route of design CDW phases with novel physical properties.

## ACKNOWLEDGMENTS

The authors gratefully acknowledge the support of NSFC (Grants Nos. 11974269, 51728203, 51621063, 51601140, 51701149, 51671155), the support by 111 project 2.0 (Grant No. BP2018008), the National Science Basic Research Plan in the Shaanxi Province of China (2018JM5168) and Innovation Capability Support Program of Shaanxi (Nos. 2018PT-28, 2017KTPT-04). J.D. also thanks the Fundamental Research Funds for the Central Universities. J.Z.L. acknowledges the support from ARC discovery projects (DP180101744) and HPC from National Computational Infrastructure from Australia. This work is also supported by State Key Laboratory for Mechanical Behavior of Materials and HPC platform of Xi'an Jiaotong University.

The authors would also like to thank Mr. F. Yang and Dr. X. D. Zhang at Network Information Center of Xi'an Jiaotong University for support of HPC platform.

Fig. 1

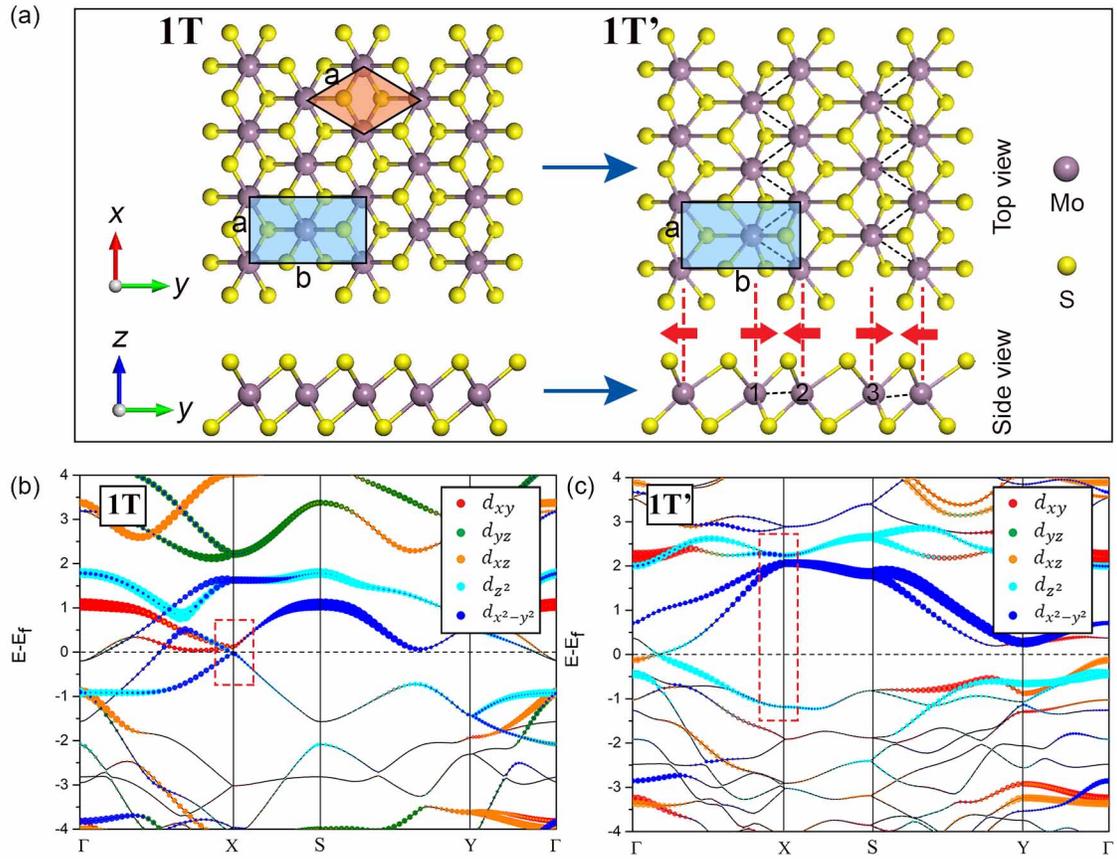

**Fig. 1 Crystal and electron structures of 1T and 1T'-MoS$_2$.** (a) Top and side view crystal structures of 1T and 1T' phase. The orange rhombus is the primitive unit of 1T-MoS$_2$, while the blue rectangle is the primitive unit of 1T' phase. The primitive unit of 1T' phase corresponds to 1× √3 supercell of 1T phase. The black dash line represents dimerization between two adjacent lines of Mo atoms. (b) and (c) are band structures of 1T and 1T' phase, respectively. The $d$ orbitals of Mo atom are mapped with different colors: $d_{xy}$, red; $d_{yz}$, green; $d_{xz}$, orange; $d_{z^2}$, cyan; $d_{x^2-y^2}$, blue.

**Fig. 2**

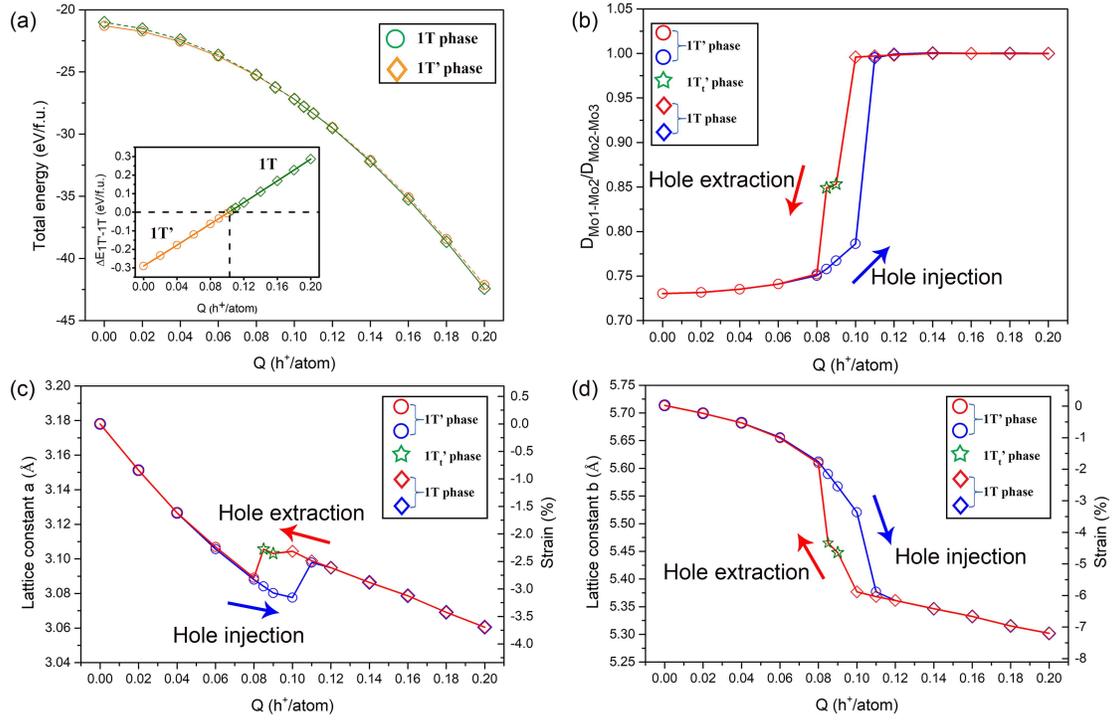

**Fig. 2 Hole injection and extraction induced reversable phase transition.** (a) The total energy of 1T and 1T' phase with different hole injection. The inset is the total energy difference between the 1T' and 1T phase as a function of hole doping/injection. (b) Interatomic distance ration $D_{Mo1-Mo2}/D_{Mo2-Mo3}$ changes with hole injection/extraction. This ratio can be used quantitative indication of different phases. The high symmetry 1T phase has this ratio ~ 1.0. The 1T' phase has this ratio of ~ 0.75. There is new phase $1T_t'$ (see main text for details) with this ratio ~ 0.85. (c) The variation of constant *a* along *x*-direction upon hole injection/extraction. (d) The variation of lattice constant *b* along *y*-direction upon hole injection/extraction. The blue and red arrows are used to denote hole injection and extraction processes, respectively. The blue and red diamond symbols represent 1T phase in hole injection and extraction. The blue and red round symbols represent 1T' phase in hole injection and extraction. The green star represents a new phase $1T_t'$ (see main text for details).

**Fig. 3**

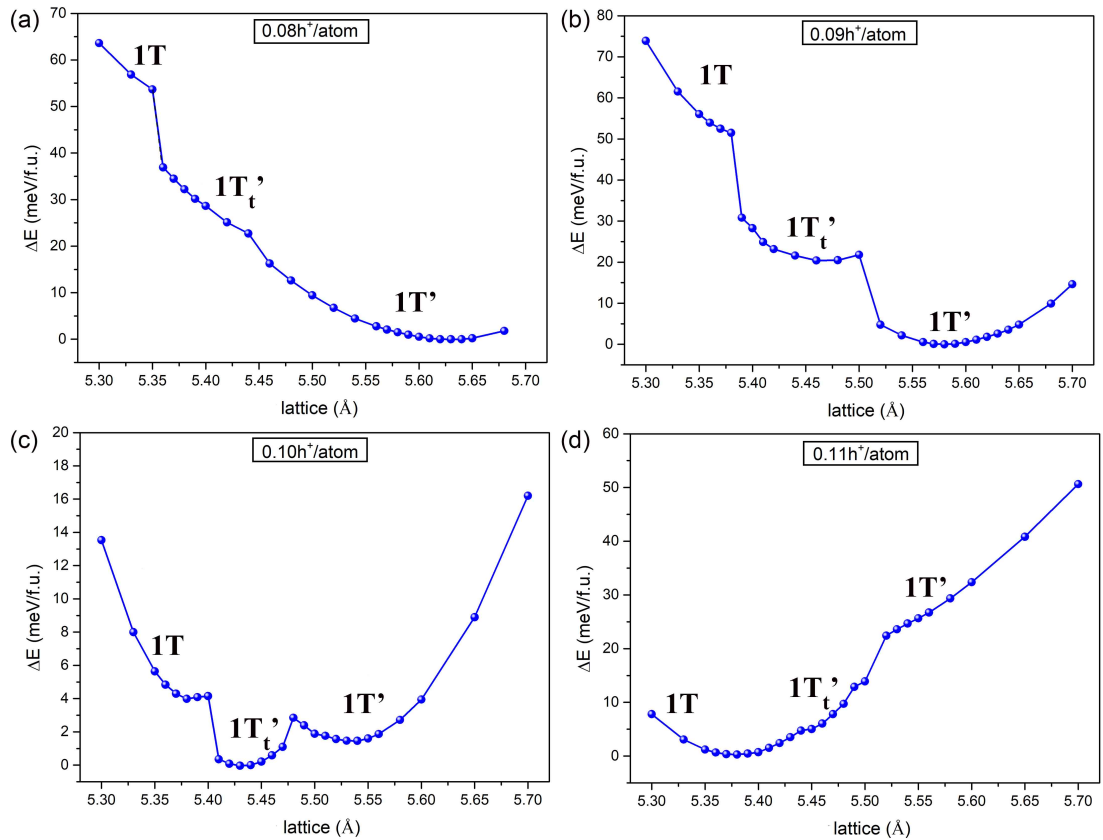

**Fig. 3 Relative total energy of MoS$_2$ versus lattice constant *b* along *y*-direction under different hole injection. (a)** Q = 0.08h$^+$/atom, **(b)** Q = 0.09h$^+$/atom, **(c)** Q = 0.10h$^+$/atom, and **(d)** Q = 0.11h$^+$/atom. There are three different phases of MoS$_2$. The new phase is termed as 1T$_t$', owing to its apparent structure similarity to 1T'. Hole injection alters the energetic orders of the three different phases and thus their relative stability.

**Fig. 4**

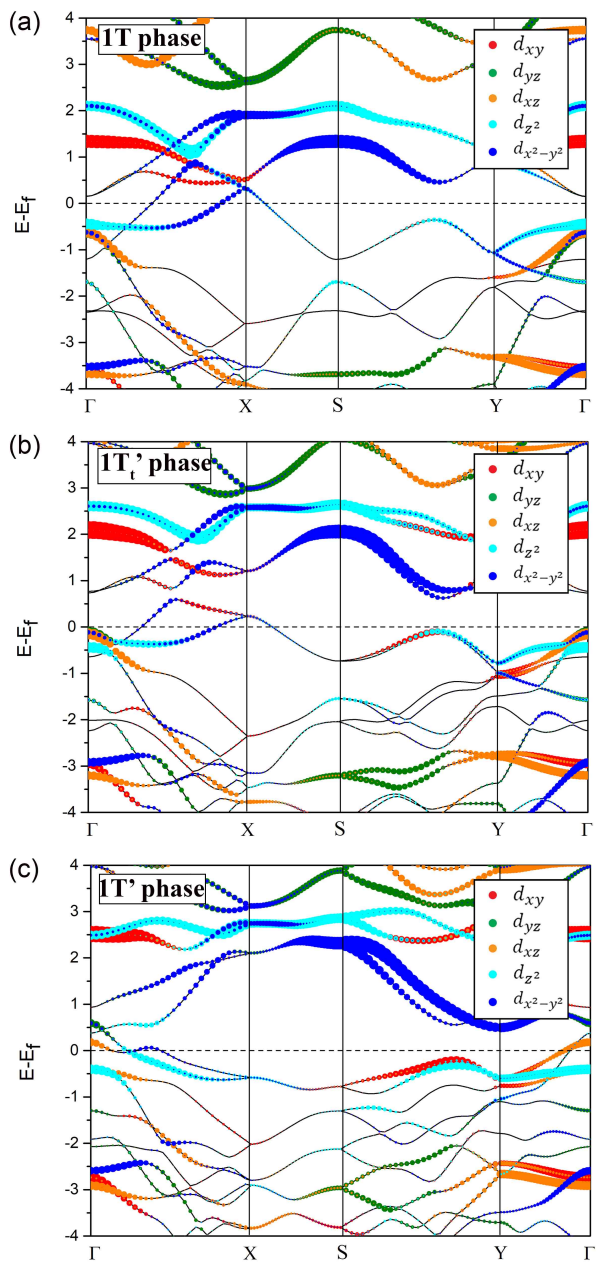

**Fig. 4 Electron band structures of 1T, 1T$_t$', and 1T' phases under Q = 0.10h$^+$/atom.** (a) 1T phase. (b) 1T$_t$' and (c) 1T'. The $d$ orbitals of Mo atom are mapped with different colors: $d_{xy}$, red; $d_{yz}$, green; $d_{xz}$, orange; $d_{z^2}$, cyan; $d_{x^2-y^2}$, blue. 1T$_t$' might be regarded as the transition phase between 1T and 1T' because its band structure shares similarity to both 1T and 1T' phases.

Fig. 5

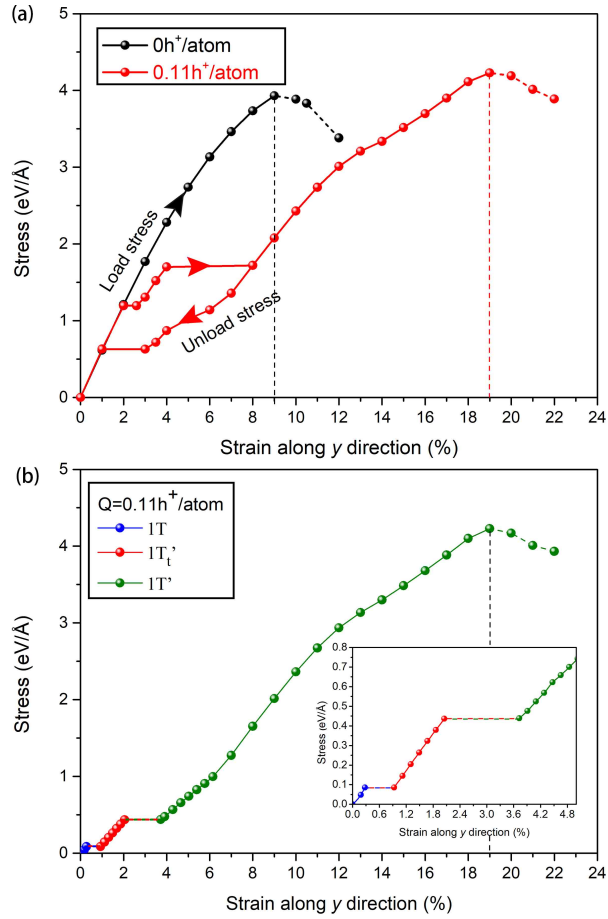

**Fig. 5 Superelasticity of MoS$_2$ under Q = 0.11h$^+$/atom hole doping.** (a) The stress-strain curves of MoS$_2$ under zero or 0.11h$^+$/atom hole injection and under strain-controlled loading condition. The charge neutral MoS$_2$ (1T' phase) shows a maximum strength of ~ 3.93eV/Å at strain ~ 9%, beyond which it starts to fail. For the charged MoS$_2$ under Q = 0.11h$^+$/atom (1T phase), there are two plateaus. The characteristic plastic-like plateau corresponds to mechanical loading induced phase transition. The two plateaus at ~2% and ~4% correspond to phase transition from 1T to 1T$_t$' and 1T$_t$' to 1T', respectively. The yield strain of the charged MoS$_2$ is ~19%, a two-fold increase compared with charge neutral 1T' MoS$_2$. (b) The stress-strain curves of MoS$_2$ under 0.11h$^+$/atom hole injection and under stress-controlled loading condition. Two plateaus can be observed, from 0.3% to 0.9%, and from 2% to 3.7%, corresponding to 1T to 1T$_t$' and 1T$_t$' to 1T' phase transition. The mechanical yield point is at ~19% strain, the same to the one of strain-control case in (a).

**Fig. 6**

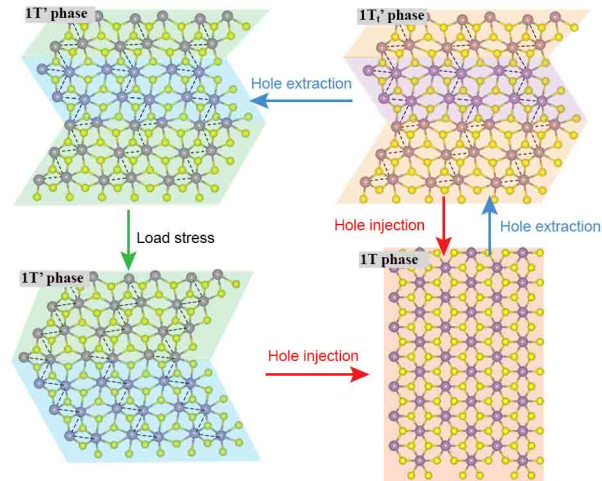

**Fig. 6 A diagram to demonstrate potential multi- shape memory effect in MoS$_2$.** The multi-domain 1T' is the permanent shape. Mechanical loads could be used to change the domain structures in 1T' and thus fix the 1T' MoS$_2$ into various new temporary shapes (the so-called shape fixing step). In the shape-recovery step, firstly holes are injected. The 1T' MoS$_2$ with temporary shapes will transform to 1T phase. The domain structure disappears since 1T is a high symmetry phase. With the subsequent hole extraction, the 1T phase will transform back to the 1T' phase with domain structures, *i.e.*, the permanent shape. Note that in this step, the new phase 1T$_t$' will appear and enable another temporary shape. This could lead to the multi- shape memory effect. Note this diagram is constructed by analogy to shape memory alloy. In this sense, the 1T phase corresponds to the austenite, whereas the 1T' corresponds to the martensite.

# Support information

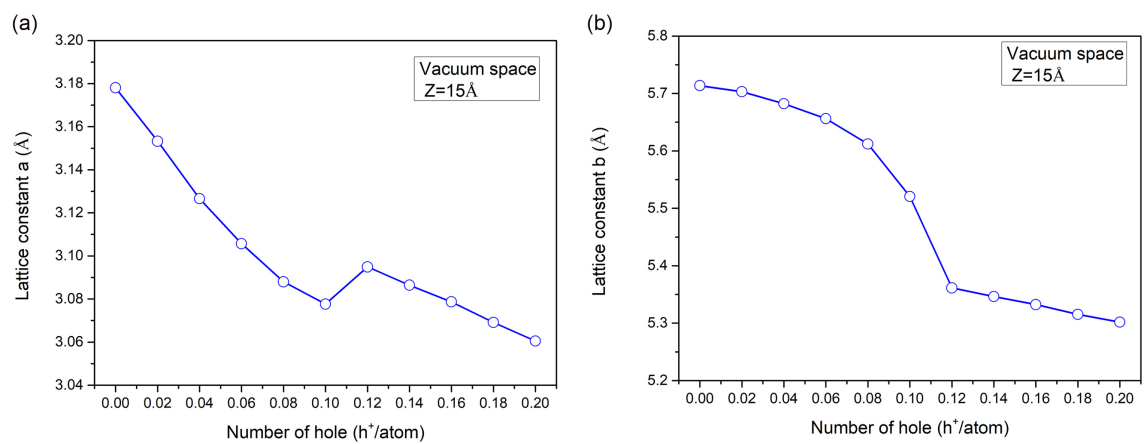

**Fig. S1** Lattice constants of 1T'-MoS$_2$ as function of hole injection. The supercell has a size of 15 Å in z-direction. A sudden change can be observed close to 0.10 h$^+$/atom. This represents 1T' to 1T phase transition.

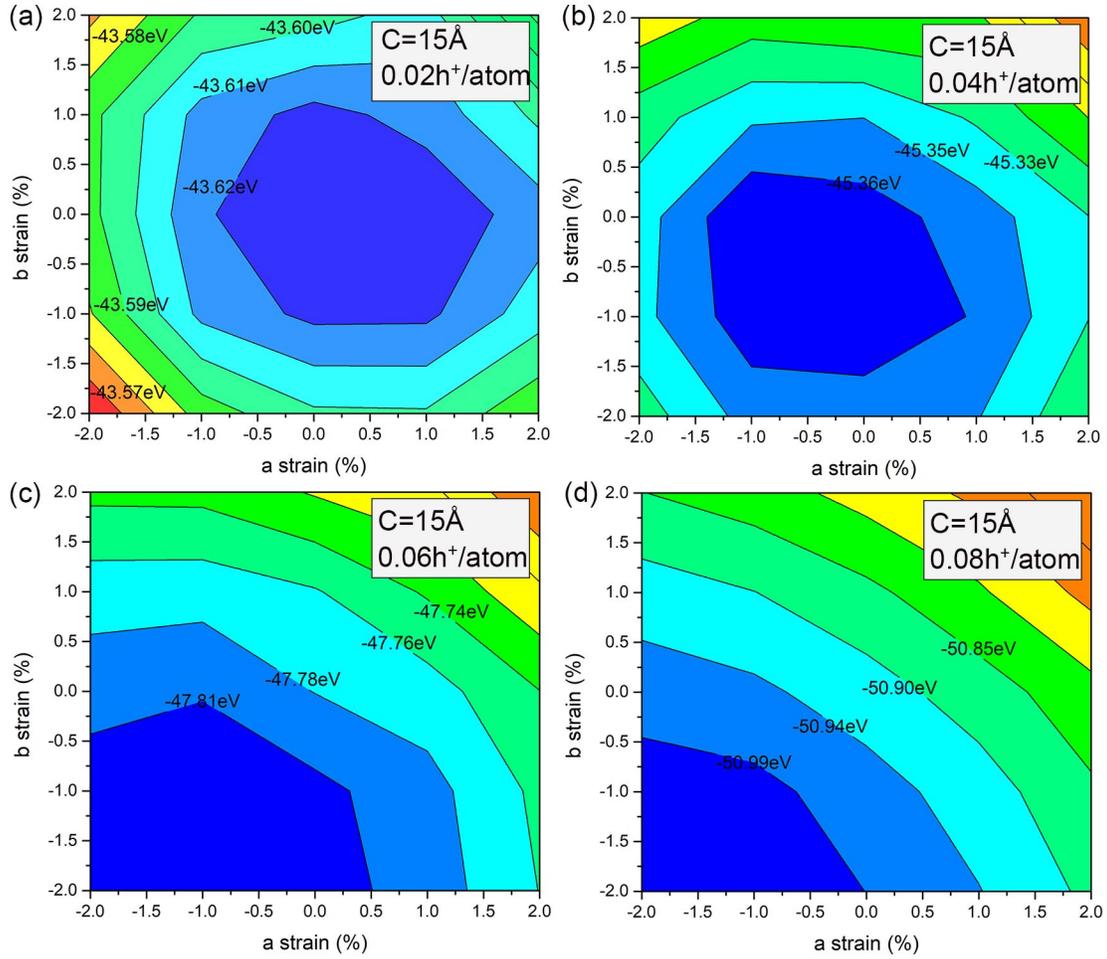

**Fig. S2** The total energy of hole doped MoS2 as a function of fixed in-plane strain $\varepsilon_x$ and $\varepsilon_y$. The hole doping level is (a) 0.02, (b) 0.04, (c) 0.06 and (d) 0.08h$^+$/atom, respectively. The supercell used in our calculation has a size of 15Å in z-direction. The lattice constants a and b are fixed in a range from -2% to 2% compared with neutral state. All energies are corrected by using the method proposed by Reed. It can be clearly seen that both lattice constants *a* and *b* decrease as the positive charge increasing.

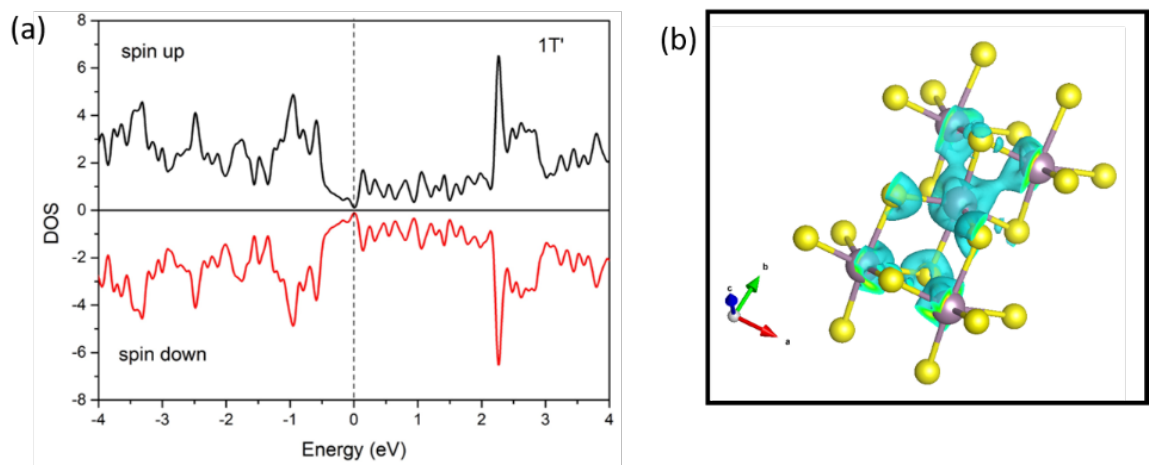

**Fig. S3 Electron structure of 1T' MoS$_2$. (a)** DOS of 1T'-MoS$_2$. **(b)** Partial electron density distribution ranges -1.5eV to Fermi level, clearly showing the Mo-Mo dimer bonds

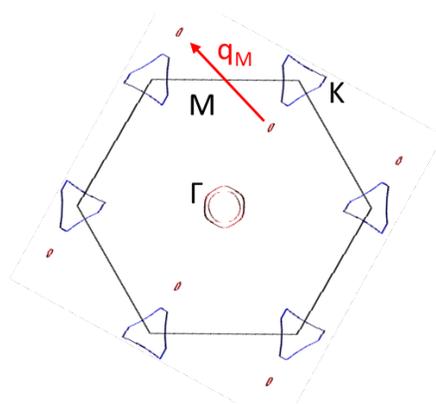

**Fig. S4** Fermi surface of monolayer 1T-MoS$_2$. The red and blue pockets stand for electron and hole Fermi surface, respectively. The nesting vector **q$_M$** connects the electron pockets (solid red line), demonstrating the instability of 1T-MoS$_2$.

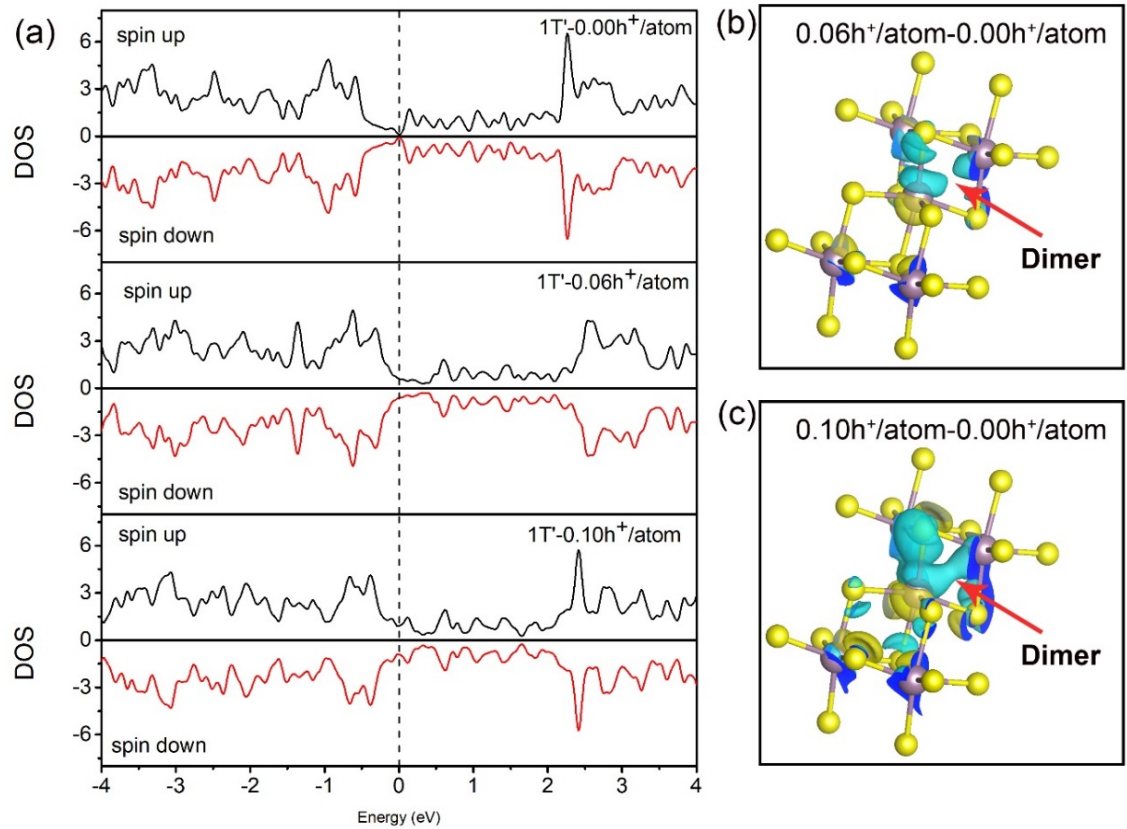

**Fig. S5 Electron structures of 1T'-MoS$_2$ with hole injection. (a)** Electronic density of states (DOS) under different hole injection. **(b)** and **(c)** show excess charge distribution for the 0.06 h$^+$/atom and 0.10 h$^+$/atom cases, respectively. The blue color represents electron depletion.

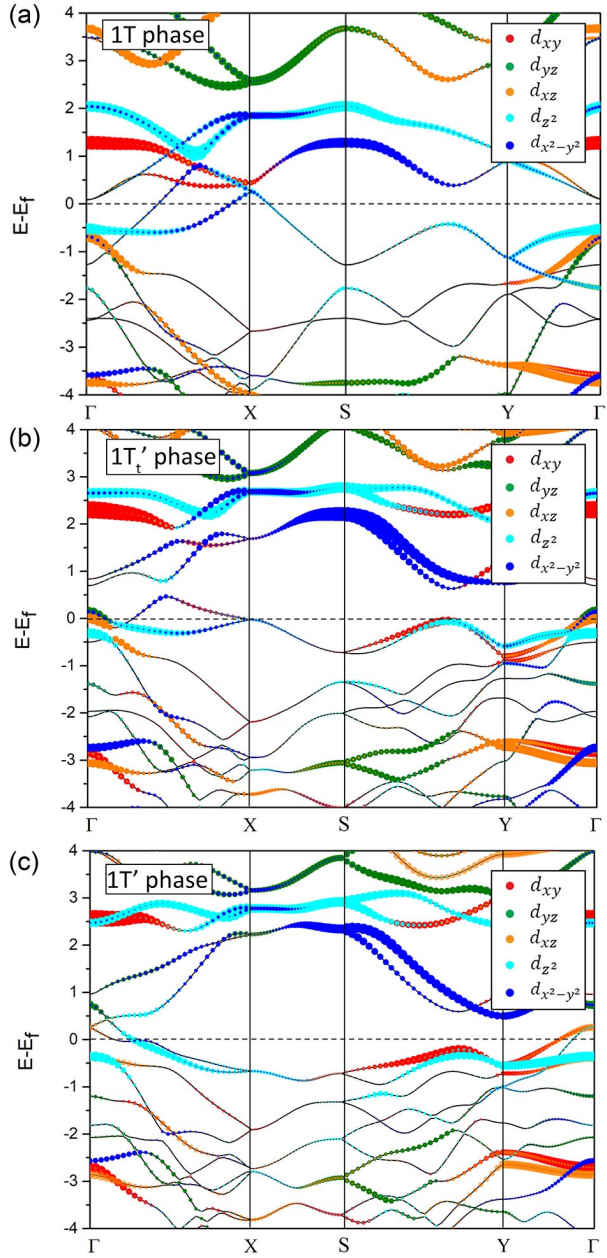

**Fig. S6 Electron band structures of 1T, 1T$_t$' and 1T' under Q = 0.09h$^+$/atom.** The *d* orbitals of Mo atom are mapped with different colors: $d_{xy}$, red; $d_{yz}$, green; $d_{xz}$, orange; $d_{z^2}$, cyan; $d_{x^2-y^2}$, blue.

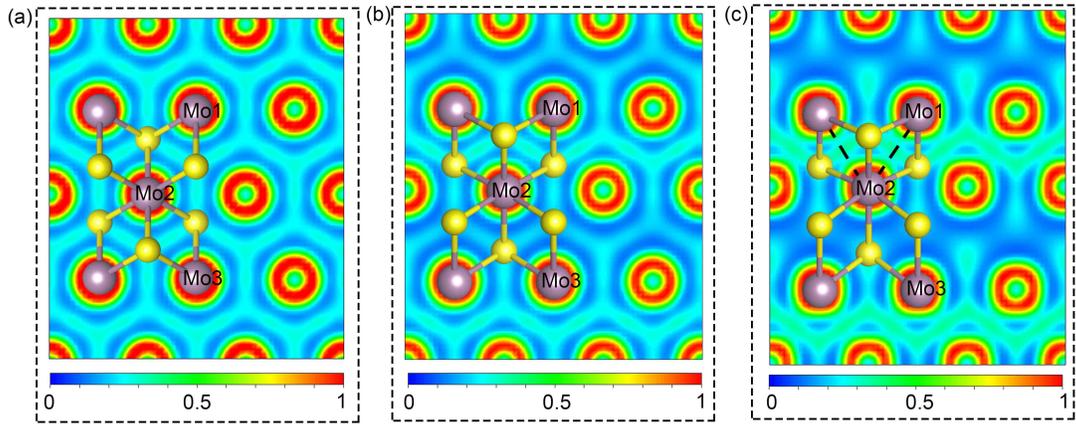

**Fig. S7 Electron local function (ELF) of 1T, 1T$_t$' and 1T' phase with Q=0.09h$^+$/atom for** (a) 1T, (b) 1T$_t$', and (c) 1T'. The 1T$_t$' has ELF results similar to that of 1T'.

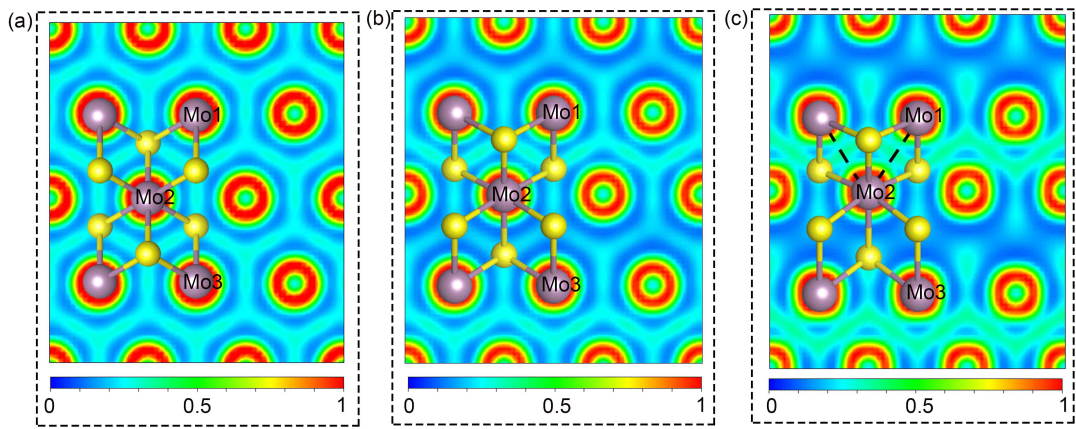

**Fig. S8 Electron local function (ELF) of 1T, 1T$_t$' and 1T' phase with Q=0.10h$^+$/atom.** The 1T$_t$' has ELF results similar to that of 1T instead of 1T'.

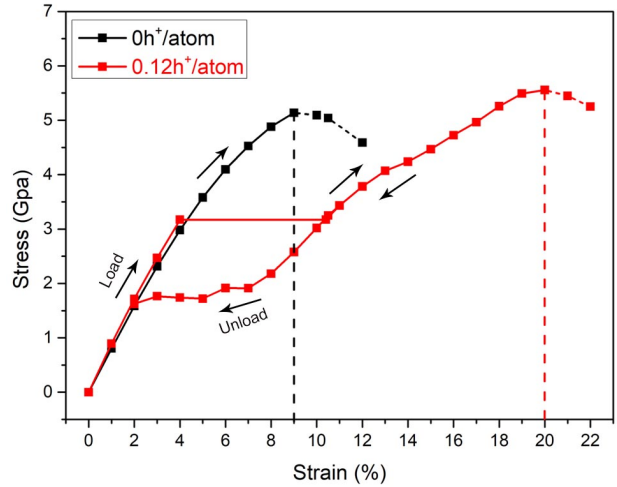

**Fig. S9 Stress−strain relation of MoS$_2$ (uniaxial tension in *y*-direction) under zero or 0.12h$^+$/atom hole injection.** The plateau corresponds to the 1T to 1T' phase transition. Only one plateau is observed in contrast to Fig. 5.

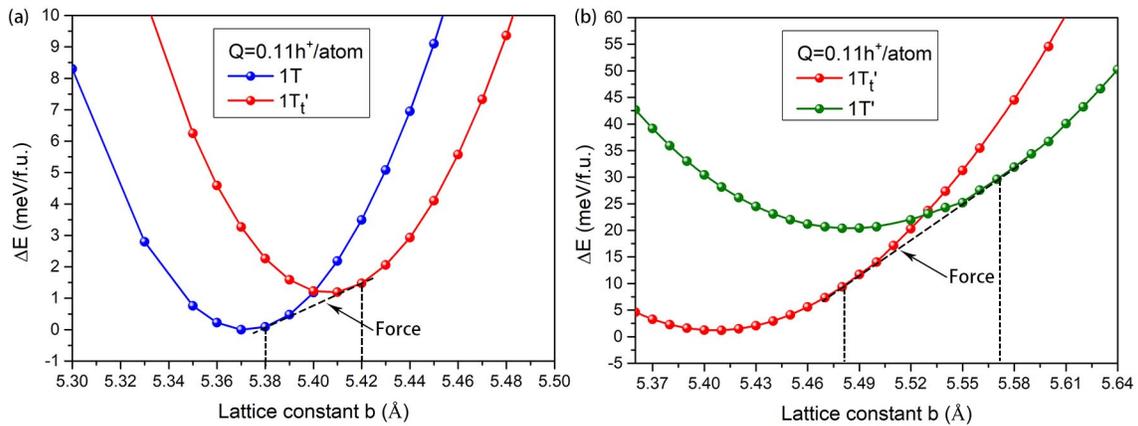

**Fig. S10 Relative total energy of MoS$_2$ versus lattice constant along *y* direction under Q=0.11h$^+$/atom.** (a) The energy change with the change of lattice constant *b*. (b) The energy change with the change of lattice constant *b*. Note that Helmholtz free energy equals to total energy at zero Kelvin. The common tangent line is used to determine the critical stress for phase transition. This method is used to construct the stress-strain relation in Fig. 5b, *i.e.*, superelasticity under stress-control condition.